\documentclass[twocolumn,showpacs,preprintnumbers,amsmath,amssymb,superscriptaddress,prl]{revtex4}

\usepackage{graphicx}
\usepackage{dcolumn}
\usepackage{bm}

\begin{document}


\title{Quantum tunneling of magnetization in single molecular magnets coupled to ferromagnetic reservoirs}

\author{Maciej Misiorny}
\affiliation{%
Department of Physics, Adam Mickiewicz University, 61-614
Pozna\'{n}, Poland
}%
\author{J\'{o}zef Barna\'{s}}%
 \email{barnas@amu.edu.pl}
\affiliation{%
Department of Physics, Adam Mickiewicz University, 61-614
Pozna\'{n}, Poland
}%
\affiliation{%
Institute of Molecular Physics, Polish Academy of Sciences, 60-179
Pozna\'{n}, Poland
}%

\date{\today}

\begin{abstract}
The role of spin polarized reservoirs in quantum tunneling of
magnetization and relaxation processes in a single molecular magnet
(SMM) is investigated theoretically. The SMM is exchange-coupled to
the reservoirs and also subjected to a magnetic field varying in
time, which enables the quantum tunneling of magnetization. The spin
relaxation times are calculated from the Fermi golden rule. The
exchange interaction of SMM and electrons in the leads is shown to
affect the spin reversal due to quantum tunneling of magnetization.
It is shown that the switching is associated with transfer of a
certain charge between the leads.
\end{abstract}

\maketitle

{\it Introduction} Single molecular magnets (SMMs) form a class of
systems whose permanent magnetic moments stem from their molecular
structure~\cite{Sessoli-etalNature365/93,Gatteschi-SessoliAngewChem42/03}.
Generally, SMMs are characterized by a large ground state spin
number $S$ and a relatively large uniaxial-type magnetic anisotropy.
As a result, an energy barrier appears for switching SMM's spin
between the two stable spin states $|\pm S\rangle$. At higher
temperatures SMMs  behave like paramagnetic or superparamagnetic
particles with a large magnetic moment. When temperature is lowered,
the thermal energy is not sufficient to reverse spin orientation of
the molecule.

It has been recently predicted that the molecule's spin can be
reversed by a spin current~\cite{Misiorny06,Elste06}. This method
allows magnetic switching without any external magnetic field and is
of great interest from the point of view of future applications in
spintronics and information
technology~\cite{Joachim-etalNature408/00, Timm-ElstePRB73/06}.
Another way to switch spin of the molecule relies on the phenomenon
of quantum tunneling of magnetization
(QTM)~\cite{Chudnovsky-TejadaMQTbook}, which occurs in a magnetic
field varying (linearly) in time. The quantum tunneling takes then
place between the energetically matched levels on the opposite sides
of the barrier, and leads to successive steps observed in hysteresis
loops~\cite{Thomas-etalNature383/96}.

\begin{figure}
    \includegraphics[width=0.95\columnwidth]{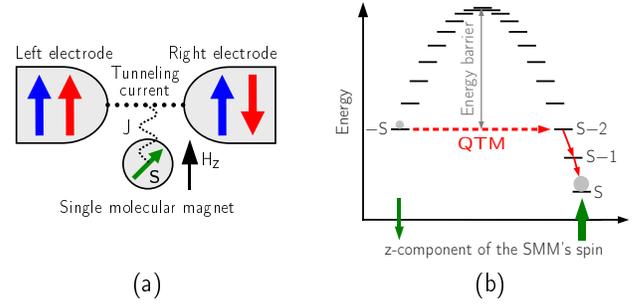}
    \caption{\label{Fig1} (color online) (a) Schema of the system under
    consideration for parallel (black solid arrows) and antiparallel (grey
    solid arrows) magnetic configurations. (b) Energy levels corresponding to different spin
    states of the SMM. The grey dots represent the occupancy of the molecular spin
    states when the third resonant field is reached.}
    \end{figure}

An important problem associated with the phenomenon of QTM is the
spin relaxation. Such a relaxation can take place for instance {\it
via} spin-orbit or spin-spin couplings. In this paper we consider
spin switching {\it via} QTM, associated with the spin relaxation
due to coupling of the molecule with two reservoirs of spin
polarized electrons. More specifically, the system under
consideration consists of a SMM located in a junction, and separated
from two ferromagnetic leads by tunnel barriers, see
fig.~\ref{Fig1}(a).  For simplicity, we take into account only
collinear (parallel and antiparallel) magnetic configurations of the
leads. Moreover, the magnetic easy axis of the molecule is parallel
to the magnetic moments of the leads. To allow the QTM, an external
time-dependent magnetic field is applied to the molecule along its
easy axis, see fig.~\ref{Fig1}(b). We assume that the field has no
influence on the magnetic configuration of the system, which can be
controlled by different means (exchange anisotropy, for instance).
In addition, we assume a significantly weaker magnetic field
perpendicular to the magnetic easy axis.

{\it Theoretical description} In a recent paper~\cite{Misiorny06} we
considered the situation when a bias voltage was applied between the
two external ferromagnetic leads, and the spin reversal was due to
the associated spin current. Here, we consider a different
situation, i.e. the system is unbiased but instead an external
magnetic field is applied to switch the molecule owing to the QTM
phenomenon. The switching is accompanied by a current impulse
between the leads, which is a consequence of the spin relaxation
processes. This, in turn, is associated with transfer of a certain
charge between the leads.

The simplest model Hamiltonian describing the SMM takes the
form~\cite{vanHammen-SutoEurphysLett1/86,
Dobrovitsky-ZvezdinEurophysLett38/97,
Wernsdorfer-SessoliScience284/99}
\begin{equation}
\mathcal{H}_{S\! M\!
M}=\mathcal{H}_{0}+\mathcal{H}_{QT\!M}+\mathcal{H}_I.
\end{equation}
The first term includes the uniaxial anisotropy and the Zeeman
energy of the molecule in a field along the anisotropy axis (the
field varies linearly in time),
    \begin{equation}\label{eq:Ham0}
    \mathcal{H}_0 = -DS_z^2 + g\mu_B S_z H_z,
    \end{equation}
where $S_z$ is the $z$-component of the spin operator, $D$ is the
uniaxial magnetic anisotropy constant, $g$ is the Land\'{e} factor,
 and $\mu_B$ stands for the Bohr magneton.

The next term, $\mathcal{H}_{QT\!M}$, of the Hamiltonian is involved
in the phenomenon of QTM, and includes the transverse anisotropy and
the Zeeman energy of the molecule in a field perpendicular to the
anisotropy axis,
    \begin{align}\label{eq:HamQTM}
    \mathcal{H}_{QT\!M}& = E(S_x^2-S_y^2)+C(S_+^4+S_-^4)
    \nonumber\\
    & + g\mu_B (S_x H_x+S_y
    H_y),
    \end{align}
where $S_x$ and $S_y$ are the transverse components of the spin
operator, $S_\pm=S_x\pm\imath S_y$, whereas $E$ and $C$ are the
transverse magnetic anisotropy constants. The perpendicular magnetic
field is much weaker than the longitudinal component $H_z$, and is
assumed to grow in time with the same time scale as $H_z$, i.e.
$H_x=0.1H_z$ ($H_y=0$ for simplicity).

The last term of the Hamiltonian, $\mathcal{H}_{I}$, describes
exchange interaction between electrons of both leads and the
SMM~\cite{AppelbaumPRL17/66, AppelbaumPR154/67, KimPRL92/04},
    \begin{equation}
    \label{eq:HamTunel}
    \mathcal{H}_I = \frac{1}{2}\sum_{q,q^\prime}\sum_{{\bf k}{\bf k}'\alpha\beta}
    \frac{J_{q,q^\prime}}{\sqrt{N_{q}\,N_{q^\prime}}}\:
    \bm{\sigma}_{\alpha\beta}\cdot\mathbf{S}\:
    a_{{\bf k}\alpha}^{q\dag}
    a_{{\bf k}'\beta}^{q^\prime}\
    \: +\:  \textrm{H.c.}.
    \end{equation}
Here, $\mathbf{S}$ is the SMM's spin operator,
$\bm{\sigma}=(\sigma^x,\sigma^y,\sigma^z)$ is the Pauli spin
operator for conduction electrons, $a^q_{{\bf k}\alpha}$
($a^{q\dag}_{{\bf k}\alpha}$) are the annihilation (creation)
operators of electrons in the left ($q=L$) or right ($q=R$)
electrodes, ${\bf k}\ ({\bf k'})$ denotes a wave vector, and
$\alpha\ (\beta)$ is the electron spin index. In
eq.~(\ref{eq:HamTunel}) $J_{q,q^\prime}$ is the relevant exchange
parameter, assumed to be independent of energy and polarization of
the leads. In a general case  $J_{L,L}\ne J_{R,R}\ne
J_{L,R}=J_{R,L}$. In the following, however, we assume the
symmetrical situation, where $J_{L,L}=J_{R,R}=J_{L,R}=J_{R,L}\equiv
J$. Owing to the proper normalization, $J$ is also independent of
the electodes' size. Finally, in eq.~(\ref{eq:HamTunel}) $N_q$
($q=L,R$) denotes the number of elementary cells in the $q$-th
electrode. Since we consider only the unbiased situation, we have
omitted the direct tunneling between the leads (i.e., tunneling
without exchange interaction with the molecule).

First, we present some general equations describing spin relaxation
of the molecule, and start from the transition rates
$\gamma_m^{>(<)}=\gamma_m^{L\! R\,>(<)}+\gamma_m^{R\! L\,>(<)}
+\gamma_m^{L\! L\,>(<)}+\gamma_m^{R\! R\,>(<)}$, at which the
molecule's spin changes from $|m\rangle$ to the upper (lower) state
$|m\pm 1\rangle$ due to coupling to the leads. Here,
$\gamma_m^{L\!R\,>(<)}$ is the transition rate of electrons
tunneling from the left to right electrode, whereas
$\gamma_m^{R\!L\,>(<)}$ corresponds to electrons tunneling the other
way. Similar meaning have the rates $\gamma_m^{L\!L\,>(<)}$ and
$\gamma_m^{R\!R\,>(<)}$. They refer, however, to electrons which
when interacting with the molecule change its magnetic state, but
remain in the same electrode. All these transition rates can be
calculated from the Fermi golden rule, and for instance
    \begin{equation}\label{eq:TransitionRatesFGR}
    \gamma_m^{L\!R\,>(<)}=\sum_{{\bf k} \alpha}\sum_{{\bf k'} \beta}
    W^{L{\bf k}\alpha m}_{R{\bf k}'\beta m\pm1}\: f(\epsilon_{{\bf k}\alpha}^L)\left[1-f(\epsilon_{{\bf
    k}'\beta}^R)\right],
    \end{equation}
where $f(\epsilon)$ is the Fermi--Dirac distribution, and
$\epsilon_{{\bf k}\alpha}^q$ ($q=L,R$) is the energy of electron
states in the leads. The probability of electron transitions from
the initial state $i\leftrightarrow\{L{\bf k}\alpha m\}$ to the
final one $j\leftrightarrow\{R{\bf k}'\beta m\pm1\}$ is given by
$W^{i}_{j}=(2\pi/\hbar )|\langle
j|\mathcal{H}_I|i\rangle|^2\delta(E_j-E_i)$, where
$E_i=\epsilon_{{\bf k}\alpha}^L+E_m$, and $E_m$ is the energy of the
unperturbed molecular spin state $|m\rangle$. Energy $E_j$ of the
final state is given by a similar expression. Finally, similar
formulae also hold for the transition rates $\gamma_m^{R\!L\,>(<)}$,
$\gamma_m^{L\!L\,>(<)}$ and $\gamma_m^{R\!R\,>(<)}$.

The final expressions for the transition rates $\gamma_m^{>(<)}$ are
given by the formulae
    \begin{align}\label{eq:TransitionRates}
    \gamma_m^{>(<)}&=
    \frac{2\pi}{\hbar} |J|^2 A_\pm(m)
    \nonumber \\
    &\hspace{0.4cm}\times
    \Big\{D_\uparrow^L D_\downarrow^R+D_\downarrow^L
    D_\uparrow^R+D_\uparrow^L D_\downarrow^L+D_\uparrow^R D_\downarrow^R\Big\}
    \nonumber \\
    &\hspace{0.4cm}\times
    \zeta\Big(\pm D(2m+1)\mp g\mu_B H_z\Big),
    \end{align}
where $D_\sigma^q$ is the density of states (DOS) at the Fermi level
in the $q$-th electrode for spin $\sigma$, $A_\pm (m)$ is defined as
$A_\pm (m)=S(S+1)-m(m\pm1)$, and
$\zeta(\epsilon)=\epsilon\big[1-\exp(-\epsilon\beta)\big]^{-1}$ with
$\beta^{-1}=k_B T$.

Since our objective is to calculate the average value of the $z$
component of SMM's spin, $\langle S_z\rangle=\sum_m mP_m$, we have
to determine the probabilities $P_m$ of finding the SMM in the spin
state $|m\rangle$. To start with, we assume the initial spin state
of the SMM to be $|-S\rangle$. Taking into account negative sign of
the gyromagnetic factor, we write $H_z=-H$. Thus, in the initial
state the magnetic field $H$ is negative. The field $H$ grows then
linearly in time and the SMM's spin undergoes transitions (due to
QTM) at relevant resonant fields from the state $|-S\rangle$ to
states $|S-M\rangle$ (consecutively for $M=0$, $M=1$, etc) on the
opposite side of the energy barrier. The probabilities of all other
states (for each $M$), i.e. of states
$|-S+1\rangle,\ldots,|S-M-1\rangle$, are then equal to zero. If the
SMM's spin tunnels to the state other than $|S\rangle$, the
interaction of the molecule with the electron reservoirs leads to
further relaxation of the spin to the final state $|S\rangle$,
fig.~\ref{Fig1}(b).

There are two time scales set respectively by the speed $c$ at which
the magnetic field is increased, $c=dH/dt$, and the relaxation
processes due to interaction with the electrodes, with the latter
scale being much shorter. This fact can be used to simplify the
problem, and the probabilities $P_m$ can be then found from the set
of master equations, separately for each field range between the
successive resonant fields at which the QTM occurs. Moreover, when
$T\ll D$, the relevant relaxation processes are those which lower
energy of the molecule (and increase the quantum number $m$ in the
case under consideration). For the $M$-th range the equations take
then the form
    \begin{equation}\label{eq:MasterEquation}
    \left\{
    \begin{aligned}
    c\, \dot{P}_{S-M}&\: =\: -\: \gamma_{S-M}^> P_{S-M},\\
    c\, \dot{P}_m&\: =\: -\: \gamma_m^> P_m\: +\: \gamma_{m-1}^> P_{m-1}, \\
    c\, \dot{P}_{S}&\: =\: \gamma_{S-1}^> P_{S-1},
    \end{aligned}
    \right.
    \end{equation}
for $m\in \langle S-M+1,S-1\rangle$, and $\dot{P}$ defined as
$\dot{P}\equiv dP/dH$. One should bear in mind that the transition
rates $\gamma_m^>$ depend on magnetic field (see
eq.~(\ref{eq:TransitionRates})). Additionally, we note that
$\gamma_m^<\approx 0$ for $m\ge 1$, which justifies the absence of
terms corresponding to the transitions to lower molecular spin
states in eqs.~(\ref{eq:MasterEquation}).

The boundary conditions for the probabilities in
eqs.~(\ref{eq:MasterEquation}) (at the resonant field $H^{(M)}$)
are:
    \begin{equation}\label{eq:BoundaryConditions}
    \left\{
    \begin{aligned}
    P_{S-M}\Big(H^{(M)}\Big)&\: =\: \widetilde{P}_{S-M},\\
    P_m\Big(H^{(M)}\Big)&\: =\: P_m^{M-1}\Big(H^{(M)}\Big) \\
    \end{aligned}
    \right.
    \end{equation}
for $m\in \langle S-M+1,S\rangle$. In
eqs.~(\ref{eq:BoundaryConditions}) $P_m^n\big(H^{(n+1)}\big)$
denotes the probability $P_m$ calculated for the field range
$\langle H^{(n)},H^{(n+1)}\rangle$ and taken at the field
$H^{(n+1)}$. Moreover, $\widetilde{P}_{S-n}$ is the probability of
the QTM to occur between the states $|-S\rangle$ and $|S-n\rangle$.
This probability can be obtained analytically with the repetitive
use of the two-level Landau-Zener model~\cite{KimPRL92/04,
 ZenerPRSL32/137},
 $\widetilde{P}_{S-n}=\big(1-\exp[-2\pi\lambda_n]\big)\prod_{l=0}^{n-1}\exp[-2\pi\lambda_l]$,
 where $\lambda_n=\Delta_n^2/\big[4\hbar(2S-n)g\mu_B c\big]$ and
 $\Delta_n$ is the splitting of the two states
 due to the $\mathcal{H}_{QT\!M}$ term in Hamiltonian.

{\it Numerical results} Numerical calculations have been obtained
for an octanuclear iron(III) oxo-hydroxo cluster of the formula
$\left[\textrm{Fe}_8\textrm{O}_2(\textrm{OH})_{12}(\textrm{tacn})_6\right]^{8+}$
(shortly $\textrm{Fe}_8$), whose total spin number is $S=10$. The
 anizotropy constants $D=0.292$ K, $E=0.046$ K,
$C=-2.9\times 10^{-5}$ K as well as the splittings $\Delta_n$ are
adopted from refs.~\cite{Wernsdorfer-SessoliScience284/99,
Rastelli-TassiPRL64/01}. We assume that $J \approx 100$ meV.
Furthermore, for both leads we assume the elementary cells are
occupied by 2 atoms contributing 2 electrons each. The density of
free electrons is assumed to be $n\approx 10^{29} m^{-3}$. The
electrodes are characterized by the polarization parameter
$P^q=(D_+^q\: -\: D_-^q)/(D_+^q\: +\: D_-^q)$, where $D_{+(-)}^q$
denotes the DOS of majority (minority) electrons in the $q$-th
electrode. The temperature of the system is assumed to be $T=0.01$
K, which is below the blocking temperature $T_B=0.36$ K of
$\textrm{Fe}_8$.

    \begin{figure}
    \includegraphics[width=0.95\columnwidth]{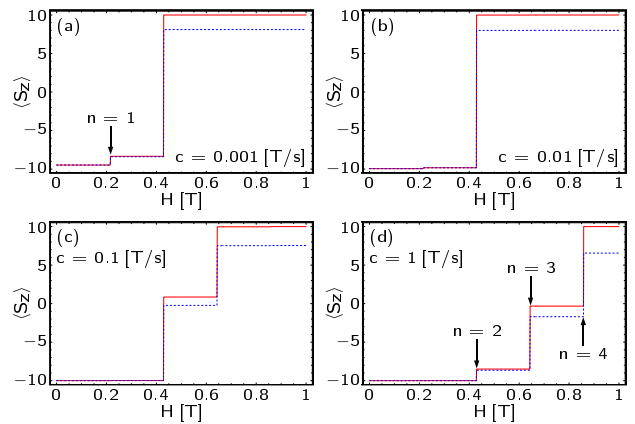}
    \caption{\label{Fig2} (color online) The average value of the SMM's spin, $\langle S_z\rangle$, as a function
    of external magnetic field $H$ for various field sweeping speeds
    $c$. The magnetic moments of the electrodes are in the {\it
    parallel}
    configuration and $P^L=P^R=0.5$. Solid lines represent the case
    of the SMM's spin coupled to electrons, whereas
    dashed ones describe the system without the exchange interaction between the
    molecule and electrons (and in the absence of other spin relaxation processes).
    }
    \end{figure}

Let us  consider first the average value of the {\it z} component of
SMM's spin, $\langle S_z\rangle$, in an external magnetic field
increasing linearly in time, fig.~\ref{Fig2}. The reversal of the
SMM's spin is due to the QTM, whose signature is the characteristic
staircase pattern. The steps occur at the resonant fields where the
QTM is allowed, and their heights are determined by the field
sweeping speed $c$ and parameters of the $\mathcal{H}_{QT\!M}$ term.
For small values of $c$ full reversal occurs already at the third
resonance field (corresponding to $n=2$). The exchange coupling with
reservoirs is essential to observe the full reversal of the spin --
after QTM the spin relaxes to the state $|S\rangle$.

Typical values of the transition times $1/\gamma_m^>$ (for the
parameters assumed) are of the order of $10^{-13}-10^{-12}$ s. The
duration of the SMM's spin reversal is thus set by the rate $c$ at
which the field $H$ is augmented, and it varies between $1$ s,
fig.~\ref{Fig2}(d), and $10^3$ s, fig~\ref{Fig2}(a). As a
consequence, the steps in figs.~\ref{Fig2} are very sharp.

Generally, the values of $\gamma_m^>$ depend on the magnetic
configuration of the system. Assume the same total density of states
in both electrodes, but generally different spin polarizations. In
the parallel (P) configuration $\gamma_m^{P\,
>}\propto 1-(P^L+P^R)^2/4$, whereas in the antiparallel (AP) one
$\gamma_m^{A\!P\,>}\propto 1-(P^L-P^R)^2/4$. If additionally we
assume now $P^L=P^R\equiv P$, the above formulae reduce to
$\gamma_m^{P\,>}\propto 1-P^2$ for the parallel configuration, while
in the antiparallel configuration $\gamma_m^{A\!P\,>}$ is
independent on the polarization $P$. The difference between these
two configurations can be clearly seen for $P=1$, which corresponds
to the situation with both electrodes being perfect halfmetallic
ferromagnets. In this limiting situation $\gamma_m^{P\,
>}\rightarrow 0$ while $\gamma_m^{A\!P\,>}$ is nonzero and independent of $P$.

The quantum tunneling phenomenon leads to another interesting
feature of the system under consideration. Generally, variation of
the $S_z$ component of the SMM's spin by one is associated with spin
reversal of one of the electrons in the leads. In the considered
situation, an electron flips its spin from $\uparrow$ to
$\downarrow$, which is accompanied by the transition of the molecule
state from $|m\rangle$ to $|m+1\rangle$. This process is continued
until the probability of finding the SMM in the spin state
$|S\rangle$ is equal to 1, which corresponds to full reversal of the
SMM's spin. Moreover, some of the spin-flip relaxation processes are
associated with a transfer of a single electron charge from one
electrode to the other. Owing to the spin asymmetry, the average
charge transferred between the leads during magnetic moment
switching becomes nonzero. As a consequence, one can expect a
current pulse ($I_r$) associated with the spin relaxation after
crossing each resonant magnetic field~\cite{Elste06a}. The general
formula for the relaxation current flowing from the left to right
side of the junction is
$I_r=e\sum_{j=1}^M\sum_{m=1}^jP_{S-m}\big(\gamma_{S-m}^{R\!L\,>}-\gamma_{S-m}^{L\!R\,>}\big)$,
where $e$ is the electron charge, and $P_{S-m}$ as well as
$\gamma_{S-m}^{R\!L(L\!R)\,>}$ depend on the field $H$. Here, $M$ is
determined from the condition $\widetilde{P}_{S-(M+1)}\approx 0$.
The time scale on which  relaxation currents vanish is of the order
of the relaxation times. For these reasons, current impulses are
very short for the parameters corresponding to fig.~\ref{Fig2} and
would not be resolved.

    \begin{figure}
    \begin{center}
    \includegraphics[width=0.95\columnwidth]{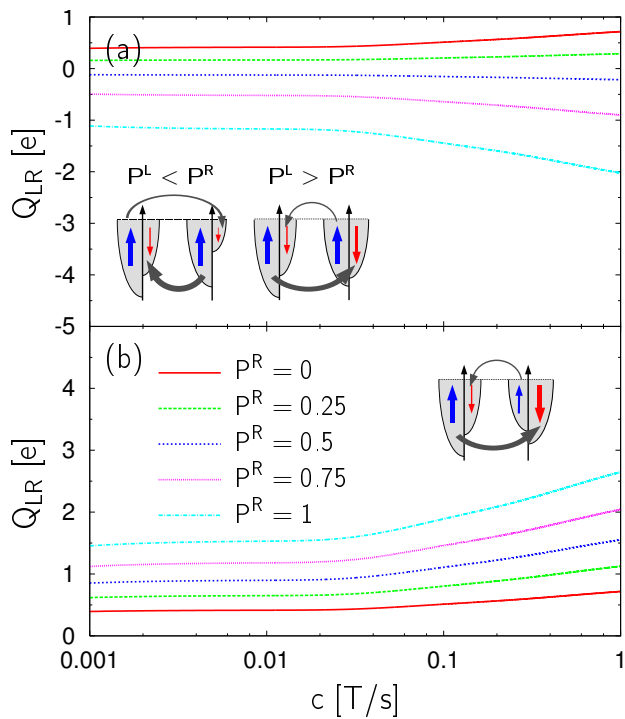}
    \caption{\label{Fig3} (color online) The average charge $Q_{L\!R}$
    transferred between the electrodes during the SMM's spin
    reversal process as a function of the field sweeping speed $c$
    for various  polarizations of the right electrode $P^R$. The
    polarization of the left electrode is $P^L=0.4$. The magnetic
    moments of the leads are kept either in the parallel (a) or
    in the antiparallel (b) configuration. The insets
    represent the DOS at the Fermi level in the leads, where the arrows
    indicate the direction of the charge transfer
    between the leads.
    }
    \end{center}
    \end{figure}

We note that the transition times only slightly change when the
field is increased, and the corresponding time scale is considerably
smaller than the time scale set by $c$. Therefore one can assume
that they are constant, which allows us to find analytic expression
for the total electronic charge transferred from the left to the
right lead during the reversal of the SMM's spin, $Q_{L\!R}=e\Gamma
\sum_{j=1}^Mj\widetilde{P}_{S-j}$, see figs.~\ref{Fig3}. $\Gamma$ is
a coefficient whose value depends on the magnetic configuration of
the leads, and $\Gamma_{A\!P(P)}=2(P^L\pm P^R)/[4-(P^L\mp P^R)^2]$,
respectively for the parallel (lower sign) and antiparallel (upper
sign) magnetic configurations.

The sign of $Q_{LR}$, which is related to the direction of average
charge flow, is determined by the magnetic configuration of the
system. In the antiparallel configuration we observe only one
direction of the average charge flow, i.e. from left to right. In
the parallel configuration, on the other hand, the average flow can
be either from left to right ($P^L>P^R$) or from right to left
($P^R>P^L$). For $P^L=P^R$ there is no resultant flow of charge. The
change of the charge transfer direction can be explained by
considering the DOS at the Fermi level in both electrodes and taking
into account that a change in the SMM's spin state is possible only
when it is associated with the flip of an electron from $\uparrow$
to $\downarrow$.

In conclusions, we have considered spin reversal of a SMM due to the
phenomenon of QTM in a time dependent magnetic field and in the
presence of two external spin-polarized leads. The molecule was
assumed to be coupled to electrons in both leads {\it via} exchange
coupling. We showed that full spin reversal is associated with a
transfer of certain charge from one lead to the other, despite no
bias voltage is applied. When external bias is applied, then new
effects may be
observed~\cite{Heersche06,Jo06,Leuenberger06,Romeike06}.

{\it Acknowledgements} This work is partly supported by funds of the
Polish Ministry of Science and Higher Education as a research
project in years 2006-2009.


\begin{thebibliography}
\expandafter\ifx\csname
natexlab\endcsname\relax\def\natexlab#1{#1}\fi
\expandafter\ifx\csname bibnamefont\endcsname\relax
  \def\bibnamefont#1{#1}\fi
\expandafter\ifx\csname bibfnamefont\endcsname\relax
  \def\bibfnamefont#1{#1}\fi
\expandafter\ifx\csname citenamefont\endcsname\relax
  \def\citenamefont#1{#1}\fi
\expandafter\ifx\csname url\endcsname\relax
  \def\url#1{\texttt{#1}}\fi
\expandafter\ifx\csname
urlprefix\endcsname\relax\def\urlprefix{URL }\fi
\providecommand{\bibinfo}[2]{#2}
\providecommand{\eprint}[2][]{\url{#2}}

\bibitem{Sessoli-etalNature365/93}
R. Sessoli, D. Gatteschi, A. Caneschi and M.A. Novak, Nature(London)
{\bf 365}, 141 (1993).

\bibitem{Gatteschi-SessoliAngewChem42/03}
D. Gatteschi and R. Sessoli, Angew. Chem. Int. Ed. {\bf 42}, 268
(2003).

\bibitem{Misiorny06}
M. Misiorny and J. Barna\'s, Phys. Rev. B {\bf 75}, 134425 (2007).

\bibitem{Elste06}
F. Elste and C. Timm, cond-mat/0611108 (to be published in Phys.
Rev. B).

\bibitem{Joachim-etalNature408/00}
C. Joachim, J.K. Gimzewski and A. Aviram, Nature {\bf 408}, 541
(2000).

\bibitem{Timm-ElstePRB73/06}
C. Timm and F. Elste, Phys. Rev. B {\bf 73}, 235304 (2006).

\bibitem{Chudnovsky-TejadaMQTbook}
E. M. Chudnovsky and J. Tejada, {\it Macroscopic Quantum Tunneling
of the Magnetic Moment} (Cambridge University Press, 1998).

\bibitem{Thomas-etalNature383/96}
L.Thomas et al., Nature(London) {\bf 383}, 145 (1996).

\bibitem{vanHammen-SutoEurphysLett1/86}
J. L. van Hemmen and A. S\"{u}t\H{o}, Europhys. Lett. {\bf 1} (10),
481 (1986).

\bibitem{Dobrovitsky-ZvezdinEurophysLett38/97}
V. V. Dobrovitsky and A. K. Zvezdin, Europhys. Lett. {\bf 38} (5),
377 (1997).

\bibitem{Wernsdorfer-SessoliScience284/99}
W. Wernsdorfer and R. Sessoli, Science {\bf 284}, 133 (1999).

\bibitem{AppelbaumPRL17/66}
J. Appelbaum, Phys. Rev. Lett. {\bf 17}, 91 (1966).

\bibitem{AppelbaumPR154/67}
J. Appelbaum, Phys. Rev. {\bf 154}, 633 (1967).

\bibitem{KimPRL92/04}
 G.-H. Kim and T.-S. Kim, Phys. Rev. Lett. {\bf 92}, 137203 (2004).

\bibitem{ZenerPRSL32/137}
C. Zener, Proc. R. Soc. London, Ser. A {\bf 137}, 696 (1932).

\bibitem{Rastelli-TassiPRL64/01}
E. Rastelli and A. Tassi, Phys. Rev. B {\bf 64}, 64410 (2001).

\bibitem{Elste06a}
F. Elste and C. Timm, Phys. Rev. B {\bf 73}, 235305 (2006).

\bibitem{Heersche06}
H.B. Heersche, Z. de Groot, J.A. Folk, and H.S.J. van der Zant et
al., Phys. Rev. Lett. {\bf 96}, 206801 (2006).

\bibitem{Jo06}
 M.-H. Jo et al., Nano Lett. {\bf 6}, 2014 (2006).

\bibitem{Leuenberger06}
M.N. Leuenberger  and E.R. Mucciolo, Phys. Rev. Lett. {\bf 97},
126601 (2006).

\bibitem{Romeike06}
C. Romeike, M.R. Wegewijs, W. Hofstetter, and H. Schoeller, Phys.
Rev. Lett. {\bf 96}, 196601 (2006).

\end{thebibliography}
\end{document}